\documentclass{llncs}

\usepackage{fullpage}
\usepackage{amsmath}
\usepackage{xspace}
\usepackage{amssymb}

\newenvironment{proofof}[1]{\noindent{\bf Proof of #1:}}{\qed}

\newcommand{\M}{{\mathcal{M}}}
\newcommand{\POVM}{\mathsf{POVM}}
\newcommand{\ket}[1]{| #1 \rangle}
\newcommand{\ketbra}[1]{| #1 \rangle \langle #1 |}

\newcommand{\Tr}{{\mathsf{Tr}}}

\newcommand{\acc}{{\mathrm{acc}}}

\newcommand{\defeq}{\stackrel{\mathsf{def}}{=}}
\newcommand{\C}{{\mathbb C}}
\newcommand{\QSC}{\mathsf{ QSC}}

\newcommand{\tp}{\tilde{p}}

\newcommand{\cK}{{\cal K}}
\newcommand{\cH}{{\cal H}}
\newcommand{\cE}{{\cal E}}
\newcommand{\cD}{{\cal D}}
\newcommand{\sS}{\mathsf{S}}
\newcommand{\sD}{\mathsf{D}}
\newcommand{\suppress}[1]{}
\newcommand{\alice}{\mathsf {Alice}}
\newcommand{\bob}{\mathsf {Bob}}

\title{New binding-concealing trade-offs for quantum string commitment}
\author{
Rahul Jain\thanks{ This work was mostly done while the author was at
U.C. Berkeley, California, USA where it was supported by an Army
Research Office (ARO), North California, grant number DAAD
19-03-1-00082. Part of the work done at U. Waterloo where it is
supported in part by ARO/NSA USA. } \email{rjain@cs.uwaterloo.edu}
\institute{School of Computer Science and Institute for Quantum
Computing, University of Waterloo, \newline Waterloo, ON, Canada, N2L 3G1.}
}

\date{}

\begin{document}
\maketitle
\begin{abstract}
{\em String commitment } schemes are similar to the well studied {\em
bit commitment} schemes in cryptography with the difference that the
committing party, say $\alice$, is supposed to commit a long string
instead of a single bit, to another party say $\bob$. Similar to bit
commitment schemes, such schemes are supposed to be {\em binding}, i.e
$\alice$ cannot change her choice after committing, and {\em
concealing} i.e. $\bob$ cannot find $\alice$'s committed string before
$\alice$ reveals it. \suppress{Strong impossibility results are known
for bit commitment schemes both in the classical and quantum settings,
for example due to Mayer~\cite{mayer:imp} and Lo and
Chau~\cite{hklo:bitcomm1,hklo:bitcomm2}. In fact for approximate
quantum bit commitment schemes, trade-offs between the {\em degrees}
of cheating of $\alice$ and $\bob$, referred to as {\em
binding-concealing} trade-offs are known as well for example due to
Spekkens and Terry~\cite{terry:bitcomm}.}  Ideal commitment schemes
are known to be impossible. Even if some degrees of cheating is
allowed, Buhrman, Christandl, Hayden, Lo and
Wehner~\cite{harry:QSC}\footnote{A short version of this paper
appeared previously in~\cite{harry:QSC1}.} have recently shown that
there are some {\em binding-concealing} trade-offs that any quantum
string commitment scheme ($\QSC$) must follow. They showed trade-offs
both in the scenario of single execution of the protocol and in the
asymptotic regime of sufficiently large number of parallel executions
of the protocol.

We present here new trade-offs in the scenario of single execution of
a $\QSC$ protocol. Our trade-offs also immediately imply the trade-off
shown by Buhrman et al. in the asymptotic regime.  We show our results
by making a central use of an important information theoretic tool
called the {\em substate theorem} due to Jain, Radhakrishnan and
Sen~\cite{jain:substate}. Our techniques are quite different from that
of~\cite{harry:QSC} and may be of independent interest.
\\ \\
\noindent
{\em Key words:}  string commitment, quantum channels, observational
divergence, relative entropy, substate theorem.
\end{abstract}

\section{Introduction}
\label{sec:intro}
Commitment schemes are powerful cryptographic primitives. In a bit
commitment scheme $\alice$, the committer is supposed to commit a bit $b
\in \{0,1\}$ to $\bob$ in such a way that after the {\em commit phase}
she cannot change her choice of the committed bit. This is referred to
as the binding property. Also at this stage $\bob$ should not be able to
figure out what the committed bit is. This is referred to as the
concealing property. Later in the {\em reveal phase} $\alice$ is supposed
to reveal the bit $b$ and convince $\bob$ that this was indeed the bit
which she committed earlier. Bit commitment schemes have been very
well studied in both the classical and quantum models since existence
of such schemes imply several interesting results in cryptography. It
has been shown that bit commitment schemes imply existence of {\em
quantum oblivious transfer}~\cite{yao:oblivious} which in turn
provides a way to do any two-party secure
computation~\cite{killian:oblivious}. They are also useful in
constructing {\em zero knowledge proofs}~\cite{goldreich:crypto} and
imply another very useful cryptographic primitive called secure {\em
coin tossing}~\cite{blum:coin}. But unfortunately strong negative
results are known about them in case $\alice$ and $\bob$ are assumed to
possess arbitrary computation power and information theoretic security
is required. In this paper we are concerned with this setting of
information theoretic security with unbounded computational resources with cheating
parties.  Classically bit commitment schemes are known to be
impossible. In the quantum setting several schemes were proposed but
later several impossibility results were
shown~\cite{mayer:imp,hklo:bitcomm1,hklo:bitcomm2,Dariano07}. Negative results
were also shown for approximate implementations of bit commitment
schemes~\cite{terry:bitcomm,Dariano07} in which trade-offs were shown for
cheating probabilities of $\alice$ and $\bob$, referred to as
binding-concealing trade-offs. Interestingly however
Kent~\cite{kent:relative} has exhibited that bit-commitment can be
achieved using relativistic constraints. However we point out that in
this work we do not keep considerations of relativity into picture and
our setting is non-relativistic.

Now suppose instead of wanting to commit a bit $b \in \{0,1\}$, $\alice$
wants to commit an entire string $x \in \{0,1\}^n $. One way to do
this might be to commit all the bits of $x$
separately. Binding-concealing trade-offs of such schemes will be
limited by the binding-concealing trade-offs allowable for bit
commitment schemes. But it is conceivable that there might exist
cleverer schemes which allow for better binding and concealing
properties? This question was originally raised by
Kent~\cite{kent:bitcomm}. Let us first begin by formally defining a quantum string
commitment protocol.  Our definition is similar to the one considered
by Buhrman et al.~\cite{harry:QSC}
\begin{definition}[Quantum string commitment]
\label{def:QSC}
Let $P = \{p_x: x \in \{0,1\}^n \}$ be a probability distribution and
let $B$ be a {\em measure of information} (we define several measures
of information later). A $(n,a,b)-B-\QSC$ protocol for $P$ is a {\em
quantum communication protocol~\cite{yao:oblivious,hklo:bitcomm2}}
between $\alice$ and $\bob$. $\alice$ gets an input $x \in \{0,1\}^n$ (chosen
according to the distribution $P$), which is supposed to be the
string to be committed. The starting joint state of the
qubits of $\alice$ and $\bob$ is some pure state. There are
no intermediate measurements during the protocol and $\bob$ has a final
checking $\POVM$ measurement $\{M_y | y \in
\{0,1\}^n\} \cup \{I - \sum_y M_y\}$ (please see Sec.~\ref{sec:prelim} for definition of
$\POVM$) to determine the value of the committed string by $\alice$ or to
detect her cheating. The protocol runs in two phases called the commit
phase followed by the reveal phase.  The following properties need to
be satisfied.
\begin{enumerate}
\item {\bf (Correctness)} Let $\alice$ and $\bob$ act honestly. Let $\rho_x$
be the state of $\bob$'s qubits
at the end of the reveal phase of the protocol when $\alice$ gets input
$x$. Then $\forall x,y ~~ \Tr M_y
\rho_x = 1$ iff $x=y$ and 0 otherwise.
\item {\bf (Concealing)} Let $\alice$ act honestly and $\bob$ be possibly
cheating. Let $\sigma_x$ be the state of $\bob$'s qubits after the commit
phase when $\alice$ gets input $x$. Then the $B$ information of the
ensemble $\cE = \{p_x, \sigma_x \}$ is at most $b$. In particular this
is also true for both $\alice$ and $\bob$ acting honestly.
\item {\bf (Binding)} Let $\bob$ act honestly and $\alice$ be possibly
cheating. Let $c \in \{0,1\}^n$ be a string in a special cheating
register $C$ with $\alice$ that she keeps independent of the rest of the registers
till the end of the commit phase. Let $\rho_c'$ be the state of $\bob$'s
qubits at the end of the reveal phase when $\alice$ has $c$ in the
cheating register. Let $\tp_c \defeq
\Tr M_c \rho_c'$. Then for all input strings $x$, $$ \sum_{c \in
\{0,1\}^n } p_c \tp_c \quad \leq \quad 2^{a-n}.$$
\end{enumerate}
\end{definition}

The idea behind the above definition is as follows.  At the end of the
reveal phase of an honest run of the protocol $\bob$ figures out $x$ from
$\rho_x$ by performing the $\POVM$ measurement $\{M_x\} \cup \{ I -
\sum_x M_x \}$. He accepts the committed string to be $x$ iff $M_x$
succeeds and this happens with probability $\Tr M_x \rho_x$. He
declares $\alice$ cheating if $I - \sum_x M_x$ succeeds. Thus due to the
first condition, at the end of an honest run of the protocol, $\bob$
accepts the committed string to be exactly the input string of $\alice$
with probability 1. The second condition above takes care of the
concealing property stating that the amount of $B$ information about
$x$ that a possibly cheating $\bob$ gets is bounded by $b$. In
bit-commitment protocols, the concealing property was quantified in
terms of the probability with which $\bob$ can guess $\alice$'s bit.
Buhrman et al.~\cite{harry:QSC} in fact do consider $\bob$'s probability of guessing
$\alice$'s input string as quantifying the concealing property. However
in the proof of their trade-off result, they consider a related notion
of information as a quantification of the concealing property.  In
this paper, we use various notions of information to quantify the
concealing property of the protocol.  The third condition guarantees
the binding property. It makes sure that if a cheating $\alice$ wants to
postpone committing or wants to change her choice at the end of the commit phase, then
she cannot succeed in making an honest $\bob$ accept her new choice with good
probability, for a lot of different strings of her choice.

A few points regarding the above definition are important to note.  We
assume that the combined state of $\alice$ and $\bob$ at the beginning
of the protocol is a pure state. Given this assumption, it can be
assumed without loss of generality (due to the arguments
of~\cite{yao:oblivious,hklo:bitcomm2}) that it remains a pure state
till the end of the protocol (in an honest run). This is because
$\alice$ and $\bob$ need not apply any intermediate measurements,
before $\bob$ applies the final checking $\POVM$ at the end of the
protocol. Our impossibility result makes a critical use of this fact
and fails to hold if the starting combined state is not a pure
state. However, there are no restrictions on the starting pure state
shared between $\alice$ and $\bob$, it could even be an entangled
state between them. The impossibility result in~\cite{harry:QSC} has
also been shown under this assumption. This assumption has also been
made in showing impossibility results for bit-commitment
schemes~\cite{mayer:imp,hklo:bitcomm1,hklo:bitcomm2}. The main reason
why these arguments do not work, both for bit commitment and string
commitment schemes, if the combined state is not a pure state is that
the {\em Local Transition Theorem} (Thm.~\ref{thm:loctrans}
mentioned later) fails to hold for mixed states. It is conceivable
that, and will be interesting to see if better $\QSC$ schemes exist
when $\alice$ and $\bob$ are forced (by some third party say) to start
in some mixed state. Please look at~\cite{Dariano07} for extension of
impossibility results for bit-commitment to a very large class of protocols. 

\subsection{Measures of information} As we will see later, the notion of information
used in the above definition is very important and therefore let us
briefly define various notions of information that we will be
concerned with in this paper. The following notion of information,
referred to as the quantum mutual information or the Holevo-$\chi$
information is one of the most commonly used.
\begin{definition}[Holevo-$\chi$ information]
Given a quantum state $\rho$, the {\em von-Neumann} entropy of $\rho$
is defined as $\sS(\rho) \defeq - \Tr \rho \log_2 \rho$. Given quantum
states $\rho, \sigma$, the {\em Kullback-Leibler divergence} or {\em
relative entropy} between them is defined as $\sS(\rho \| \sigma) \defeq
\Tr \rho (\log_2 \rho - \log_2 \sigma)$. Given an ensemble $\cE = \{p_x,
\rho_x
\}$, let $\rho \defeq \sum_x p_x \rho_x$, then its Holevo-$\chi$ information
is defined as $$\chi(\cE) \quad \defeq \quad \sum_x p_x (\sS(\rho) -
\sS(\rho_x)) \quad = \quad \sum_x p_x \sS(\rho_x \| \rho). $$
\label{def:holevo}
\end{definition}

The following notion captures the amount of information
that can be made available to the real world through measurements on
the quantum encoding of a classical random variable.

\begin{definition}[Accessible information]
Let $\cE = \{p_x,\rho_x\}$ be an ensemble and let $X$ be a classical random
variable such that $\Pr(X=x) \defeq p_x$.  Let $Y^\M$, correlated with
$X$, be the classical random variable that represents the result of a
$\POVM$ measurement $\M$ performed on $\cE$. The {\em accessible
information\/}~$I_{\acc}(\cE)$ of the ensemble~$\cE$ is then defined
to be
\begin{equation}
\label{eqn-acc}
I_{\acc}(\cE) \quad  \defeq \quad \max_{\M} I(X:Y^\M).
\end{equation}
\end{definition}

As mentioned before Buhrman et al. used $\bob$'s probability of guessing
$\alice$'s input string as the measure of concealment of the
protocol.  However in the proofs of their impossibility result, they used
the following notion of information.
\begin{definition}[$\xi$ information~\cite{harry:QSC}]
The $\xi$ information of an ensemble $\cE =
\{p_x, \rho_x \}$ is defined as
$$\xi(\cE) \quad \defeq \quad n + \log_2 \sum_x \Tr(p_x\rho^{-1/2} \rho_x)^2 $$
where $\rho = \sum_x p_x \rho_x$.
\end{definition}
Let $q_x$ be the probability that $\bob$ correctly guesses $\alice$'s input
string $x$ (with $\alice$ honest) before the start of the reveal phase.  \cite{harry:QSC}
showed that any $(n,a,b)-\QSC$ protocol with $\sum_{x \in \{0,1\}^n} q_x \leq 2^b$, is
also a $(n,a,b)-\xi-\QSC$ protocol. Hence their impossibility results for
$(n,a,b)-\xi-\QSC$ protocols implied same impossibility results for
$(n,a,b)-\QSC$ protocols with $\sum_{x \in \{0,1\}^n} q_x \leq 2^b$.

In this paper we also consider a notion of {\em divergence
information}. It is based on the following notion of distance between
two quantum states, considered by Jain, Radhakrishnan and Sen~\cite{jain:substate}.
\begin{definition}[Observational divergence~\cite{jain:substate}]
 Let $\rho, \sigma$ be two quantum states. The observational divergence between them denoted $\sD(\rho \|
\sigma)$, is defined as, $$\sD(\rho \| \sigma) \quad \defeq \quad
\max_{\mathsf{M:POVM~element}} \Tr M \rho \log_2 \frac{\Tr M \rho} {\Tr M
\sigma}.$$
\end{definition}

The definition of divergence information of an ensemble is similar to
the Holevo-$\chi$ information except the notion of distance between
quantum states used is now observational divergence instead of
 relative entropy.
\begin{definition}[Divergence information]
\label{def:divinf}
Let $\cE = \{p_x,\rho_x\}$ be an ensemble and let $\rho \defeq \sum_x
p_x \rho_x$. Its divergence information 
is defined $$\cD(\cE) \quad \defeq \quad \sum_x p_x \sD(\rho_x \| \rho).$$
\end{definition}

\subsection{Previous results} The impossibility of a strong string
commitment protocol, in which both $a,b$ are required to be 0, is
immediately implied by the impossibility of strong bit-commitment
protocols.  The question of a trade-off between $a$ and $b$ was
studied by Buhrman et al. They studied this trade-off both in the
scenario of single execution of the protocol and also in the
asymptotic regime with several parallel executions of the protocol. In
the scenario of single execution of the protocol they showed the
following result.
\begin{theorem}[\cite{harry:QSC}]
\label{thm:harrysingle}
For single execution of the protocol of a $(n,a,b)$-${\xi}$-$\QSC$,
$a + b + 5 \log_2 5 - 4 \geq n$.
\end{theorem}
This then (as argued before) implied similar trade-off for a
$(n,a,b)$-$\QSC$ with $\sum_{x \in \{0,1\}^n} q_x \leq 2^b$ (where $q_x$
be the probability that $\bob$ correctly guesses $\alice$'s input string
$x$, with $\alice$ honest, before the start of the reveal phase.)  In the
asymptotic regime they showed the following result in terms of the
Holevo-$\chi$ information.

\begin{theorem}[\cite{harry:QSC}]
\label{thm:avgasym}
Let $\Pi$ be a  $(n, *, b)-\chi-\QSC$ scheme. Let $\Pi_m$ represent
$m$ parallel executions of $\Pi$.  Let $a_m$ represent the
binding parameter of $\Pi_m$ and let $a \defeq \lim_{m \rightarrow
\infty} \frac{a_m}{m}$.  Then, $ a + b \geq  n $.
\end{theorem}
There are two reason why Thm.~\ref{thm:avgasym} may appear stronger than
Thm.~\ref{thm:harrysingle}. One because there is no additive constant
and the other because for many ensembles $\cE$, $\chi(\cE) \leq
\xi(\cE)$ as we show in Sec.~\ref{sec:separation}. In fact, as we also
show in Sec.~\ref{sec:separation}, there exists ensembles $\cE$ for
which $\xi(\cE)$ is exponentially (in $n$) larger than $\chi(\cE)$.

Along with these impossibility results Buhrman et al. interestingly
also showed that if the measure of information considered is the
accessible information, the above trade-offs no longer hold. For
example there exists a $\QSC$ scheme where $a = 4 \log_2 n + O(1)$ and
$b = 4$ when measure of information is the accessible information. This
therefore asserts that the choice of measure of information is crucial
to (im)possibility. Previously Kent~\cite{kent:bitcomm} also exhibited
trade-offs for some schemes on $\alice$'s probability of cheating and the
amount of accessible information that $\bob$ gets about the committed
string.  However he did not allow $\alice$ to be arbitrarily cheating, in
particular $\alice$ could not have started with a superposition of
strings in the input register. Therefore the schemes that he
considered were truly not $\QSC$s as we have defined them.

\subsection{Our results}
We show the following binding-concealing trade-off for $\QSC$s.

\begin{theorem}
\label{thm:avg}
For single execution of the protocol of a $(n,a,b)-\cD-\QSC$ scheme,
$$ a + b + 8 \sqrt{b + 1} + 16  \quad \geq \quad  n. $$
\end{theorem}

It was shown by Jain, Radhakrishnan and
Sen~\cite{jain:substate} that for any two states
$\rho, \sigma$, $\sD(\rho \| \sigma) \leq \sS(\rho \| \sigma) + 1$, which
implies from Defn.~\ref{def:holevo} and~\ref{def:divinf}
that for any ensemble $\cE, \cD(\cE) \leq \chi(\cE) + 1$. This
immediately gives us the following impossibility result in terms of
Holevo-$\chi$ information.
\begin{theorem}
\label{thm:avgchi}
For single execution of the protocol of a $(n,a,b)-\chi-\QSC$ scheme
$$ a + b + 8 \sqrt{b + 2} +17  \quad \geq \quad  n. $$
\end{theorem}


We also consider the notion of {\em maximum possible
divergence information} (similar to the notion of maximum possible
Holevo-$\chi$ information considered by Jain~\cite{jain:remote}) of an
{\em encoding} $E: x \mapsto \rho_x$. For a probability
distribution $\mu \defeq \{p_x\}$ over $\{0,1\}^n$, let the ensemble
$\cE_{\mu}(E) \defeq \{p_x, \rho_x \}$.  Let $\rho_{\mu} \defeq \sum_x p_x \rho_x$.

\begin{definition}(Maximum possible divergence information)
{\em Maximum possible divergence information} of an encoding $E: x \mapsto
\rho_x$ is defined as $\tilde{\cD}(E) \defeq \max_{\mu} \cD(\cE_{\mu}(E))$.
\end{definition}
We show the following theorem which states that if the maximum
possible divergence information in the qubits of $\bob$ at the end of the commit
phase is small then $\alice$ can actually cheat with good probability for
any string $x \in \{0,1\}^n$ and not just on the average.
\begin{theorem}
\label{thm:worst}
For a $\QSC$ scheme let $\sigma_x$ be as in Defn.~\ref{def:QSC} when
$\alice$ and $\bob$ act honestly in the commit phase.  If for the encoding
$E:x \mapsto \sigma_x,
\tilde{\cD}(E) \leq b$ then for all strings $c \in \{0,1\}^n$, $$ \tp_c
\quad \geq \quad 2^{-(b + 8 \sqrt{b + 1} + 16)}, $$
where $\tp_c$ (as in Defn.~\ref{def:QSC}) represents the probability of
successfully revealing string $c$ (in the cheating string) by cheating $\alice$.
\end{theorem}
Again using the fact that for all ensembles
$\sD(\rho \| \sigma) \leq \sS(\rho \| \sigma) + 1$ we immediately get the
following theorem in terms of maximum possible Holevo-$\chi$
information $\tilde{\chi}(E)$ (which is similar to maximum possible divergence
information and obtained by just replacing divergence with relative
entropy.)
\begin{theorem}
\label{thm:worstchi}
For a $\QSC$ scheme let $\sigma_x$ be as in Defn.~\ref{def:QSC} when
$\alice$ and $\bob$ act honestly in the commit phase.  If for the encoding
$E:x \mapsto \sigma_x,
\tilde{\chi}(E) \leq b$ then for all strings $c \in \{0,1\}^n$, $$ \tp_c
\quad \geq \quad 2^{-(b + 8 \sqrt{b + 2} + 17)}, $$
where $\tp_c$ (as in Defn.~\ref{def:QSC}) represents the probability of
successfully revealing string $c$ (in the cheating string) by cheating $\alice$.
\end{theorem}

Now let us now discuss some aspects of our results.

\begin{enumerate}

\item
In Thm.~\ref{thm:avgchi} the trade-off between $a$ and $b$ is similar
(up to lower order terms of $b$) to the one shown by Buhrman et
al.~\cite{harry:QSC} as in Thm.~\ref{thm:harrysingle}. However the fact
that $b$ in Thm.~\ref{thm:avgchi} represents the Holevo-$\chi$
information instead of the $\xi$-information (as in
Thm.~\ref{thm:harrysingle}) makes it significantly stronger in certain
cases as follows.  We show in Sec.~\ref{sec:separation} that for any
ensemble $\cE \defeq \{2^{-n}, \rho_x \}$, where for all $x$, $\rho_x$ commutes with
$\rho \defeq \sum_x 2^{-n}\rho_x$, we have, $\xi(\cE) \geq
\chi(\cE)$. In fact, as we also show in Sec.~\ref{sec:separation},
there exists ensembles $\cE$ for which $\xi(\cE)$ is exponentially (in $n$) larger than
$\chi(\cE)$.  Thm.~\ref{thm:avgchi} therefore becomes much stronger
than Thm.~\ref{thm:harrysingle} for ensembles where $\xi(\cE) \gg \chi(\cE)$.

\item As mentioned before, Jain, Radhakrishnan and Sen~\cite{jain:substate}
have shown that for any ensemble $\cE, \cD(\cE) \leq \chi(\cE) +
1$. However recently, Jain, Nayak and Su~\cite{JainNS08} have shown that
there exists ensembles $\cE$ such that $\chi(\cE) \gg
\cD(\cE)$ ($\chi(\cE) = \Omega(\log_2 n \cdot \cD(\cE))$ for some ensembles
$\cE$ supported on $\{0,1\}^n$). For ensembles where this holds,
Thm.~\ref{thm:avg} becomes much stronger than Thm.~\ref{thm:avgchi}.

\item As we show in Sec.~\ref{sec:proof}, our one shot result
Thm.~\ref{thm:avgchi} immediately implies the asymptotic result
Thm.~\ref{thm:avgasym} of Buhrman et al.

\item No counterparts of Thm.~\ref{thm:worst} and
Thm.~\ref{thm:worstchi} were shown by Buhrman et al. and are therefore 
completely new.

\item If $b$ is large then the cheating attack (that we present) of $\alice$ would
succeed with low probability (like $2^{-b}$). However, as we show in a
remark in Sec.~\ref{sec:proof}, in case $\alice$'s cheating attack succeeds with
low probability, she would still be able to 'reverse' her cheating
operations and reveal, with a high probability, at least some $x' \in
\{0,1\}^n$ to $\bob$. That is, with a high probability, $\alice$ will
be able to prevent herself from being detected cheating by $\bob$.

\item It is easily seen that up to lower order terms in $b$, the above trade-offs are
achieved by trivial protocols. For Thm.~\ref{thm:avg} above
consider the following protocol. $\alice$ in the concealing phase sends
the first $b$ bits of the $n$-bit string $x$. In this case $\bob$ gets to
know $b$ bits of divergence information about $x$. In the reveal phase
a cheating $\alice$ can now reveal any of the $2^{n-b}$ strings $x$
(consistent with the first $b$ bits being the ones sent) with
probability 1. Hence $a = \log_2 2^{n-b} = n-b$. For
Thm.~\ref{thm:worst} above let $\alice$ send one of the $2^b$ strings
$s \in \{0,1\}^b$ uniformly to $\bob$ representing the first $b$ bits of
$x$. The condition of Thm.~\ref{thm:worst} is satisfied. Now if in
the reveal phase she wants to commit any $x$, she can do so with
probability $2^{-b}$ (in the event that the sent $s$ is consistent
with $x$).

\end{enumerate}

In the next section we state some quantum information theoretic facts
that will be useful in the proofs of the impossibility results that we
present in Sec.~\ref{sec:proof}.

\section{Preliminaries}
\label{sec:prelim}
All logarithms in this paper are taken with base 2 unless otherwise
specified. Let $\cH, \cK$ be finite dimensional Hilbert spaces. For a
linear operator $A$ let $|A| = \sqrt{A^{\dagger}A}$ and let $\Tr A$
denote the trace of $A$.  Given a state $\rho \in \cH$ and a pure
state $\ket{\phi} \in \cH \otimes
\cK$, we call $\ket{\phi}$ a {\em purification} of $\rho$ iff
$\Tr_{\cK}
\ketbra{\phi} = \rho $. A {\em positive operator-valued
measurement $(\POVM)$ element} $M$ is a positive semi-definite operator
such that $I - M$ is also positive semi-definite, where $I$ is the
identity operator.  A $\POVM$ is defined as follows.
\begin{definition}[$\POVM$]
An $m$ valued $\POVM$ measurement $\M$ on a Hilbert space $\cH$ is a set of operators
$\{M_i, i \in [m]\}$ on $\cH$ such that $\forall i, M_i$
is positive semi-definite and $\sum_{i \in [m]} M_i = I$ where $I$ is the
identity operator on $\cH$. A classical random variable $Y^{\M}$
representing the result of the measurement $\M$ on a state $\rho$ is an $m$ valued
random variable such that $\forall i \in [m], \Pr[Y^{\M} = i ] \defeq \Tr M_i\rho$.
\end{definition}

Following fact follows easily from definition of von-Neumann entropy.
\begin{lemma}
\label{lem:sadd}
Let $\rho_1, \rho_2$ be quantum states. Then $\sS(\rho_1 \otimes \rho_2)
= \sS(\rho_1) + \sS(\rho_2)$.
\end{lemma}

We make a central use the following information-theoretic result called
the substate theorem due to Jain, Radhakrishnan, and
Sen~\cite{jain:substate}.
\begin{theorem}[Substate theorem, \cite{jain:substate}]
\label{thm:substate}
Let $\cH, \cK$ be two finite dimensional Hilbert spaces and $\dim(\cK)
\geq \dim(\cH)$. Let $\C^2$ denote the two dimensional complex Hilbert
space.  Let $\sigma, \tau$ be density matrices in $\cH$ such that
$\sD(\sigma \| \tau) < \infty$.  Let $\ket{\overline{\sigma}}$ be a
purification of $\sigma$ in $\cH \otimes \cK$. Then, for $r > 1$, there
exist pure states $\ket{\phi}, \ket{\theta} \in \cH
\otimes \cK$ and
$\ket{\overline{\tau}} \in \cH \otimes \cK \otimes \C^2$, depending
on $r$, such that $\ket{\overline{\tau}}$ is a purification of
$\tau$ and $\Tr |\ketbra{\overline{\sigma}} - \ketbra{\phi}| \leq \frac{2}{\sqrt{r}}$, where
\begin{displaymath}
\ket{\overline{\tau}} \defeq
\sqrt{\frac{r-1}{r 2^{r k}}} \, \ket{\phi}\ket{1} +
\sqrt{1 - \frac{r-1}{r 2^{r k}}} \, \ket{\theta}\ket{0}
\end{displaymath}
and $k \defeq \sD(\sigma \| \tau) +
6 \sqrt{\sD(\sigma \| \tau) + 1} + 4$.
\end{theorem}

\paragraph{Remarks:}
\begin{enumerate}
\item In the above theorem if the last qubit in
$\ket{\overline{\tau}}$ is measured in the computational basis, then
probability of obtaining 1 is $(1 - 1/r) 2^{-r k}$.
\item  Later in a proof below we will let $\sigma \defeq \rho_c$
, $\tau \defeq \rho_B$ and $\ket{\overline{\sigma}}
\defeq \ket{\phi_c}$ which will be explained later.
\end{enumerate}

Following theorem is implicit
in~~\cite{jozsa:loctrans,mayer:imp,hklo:bitcomm1,hklo:bitcomm2}
although not called explicitly by the same name.
\begin{theorem}[Local transition theorem]
\label{thm:loctrans}
Let $\rho$ be a quantum state in $\cK$. Let $\ket{\phi_1}$ and $
\ket{\phi_2}$ be two purification of $\rho$ in $\cH \otimes \cK$. Then
there is a local unitary transformation $U$ acting on $\cH$ such that
$(U \otimes I) \ket{\phi_1} = \ket{\phi_2}$.
\end{theorem}

We would also need the following theorem which follows from arguments
similar to the one in Jain~\cite{jain:remote} for a similar theorem
about relative entropy.
\begin{theorem}
\label{thm:remote}
Let $X$ be a finite set. Let $E: x \mapsto \rho_x$ be an encoding. Let
$\tilde{\cD}(E) \leq b$, then there exists a distribution
$\mu \defeq \{q_x \}$ on $X$ such that $$\forall x \in X, \quad \sD(\rho_x \| \rho) \quad \leq \quad 
b,$$ where $\rho \defeq \sum_x q_x \rho_x$.
\end{theorem}

The following theorem is shown by Helstrom~\cite{Helstrom67}.
\begin{theorem}
\label{thm:dist}
Given two quantum states $\rho$ and $\sigma$, the probability of
identifying the correct state is at most $\frac{1}{2} + \frac{\Tr |\rho - \sigma
|}{4}$, or in other words the probability of distinguishing them is at
most $\frac{\Tr |\rho - \sigma |}{2}$. 
\end{theorem}

\section{Proofs of impossibility}
\label{sec:proof}
\noindent
\begin{proofof}{Thm.~\ref{thm:avg}} Let us consider a $\QSC$ scheme
and let $\alice$ get input $x$. After an honest run of the commit phase,
let $\ket{\phi_x}$ be the combined state of $\alice$ and $\bob$ and $\rho_x$
be the state of $\bob$'s qubits.  Let $\cE = \{p_x,
\rho_x\}$. From the concealing property of the $\QSC$ it follows
$\sD(\cE) \leq b$. Let $c$ be the string in the cheating register $C$
of $\alice$. Consider a cheating run of the protocol by $\alice$ in
which she starts with the superposition $\sum_x
\sqrt{p_x}\ket{x}$ in the input register and proceeds with the rest
of the commit phase as before in the honest protocol. Let $\bob$ be
honest all throughout our arguments.  Since the input is classical and
$\alice$ can make its copy we can assume without loss of generality
that the operations of $\alice$ in the honest run are such that they
do not disturb the input register. Let $\ket{\psi}$ be the combined
state of $\alice$ and $\bob$ in this cheating run at the end of the
commit phase.  Let $A, B$ correspond to $\alice$ and $\bob$'s systems
respectively. Now it can be seen that in the cheating run, at the end
of the commit phase the qubits of $\bob$ are in the state $\rho_B
\defeq \Tr_A
\ketbra{\psi} = \sum_x p_x \rho_x$. Let $r > 1$ to be chosen
later. Let us now invoke substate theorem (Thm.~\ref{thm:substate}) by
putting $\sigma \defeq \rho_c,
\ket{\overline{\sigma}} \defeq \ket{\phi_c}$, 
$\tau \defeq \rho_B$ and $r \defeq r$. Let $\ket{\psi_c} \defeq
\ket{\overline{\tau}}$ be obtained from
Thm.~\ref{thm:substate} such that the extra single qubit register
$\C^2$ is also with $\alice$.  Since $\Tr_{A} \ketbra{\psi_c} =
\Tr_{A} \ketbra{\psi} = \rho_B$, from Local transition theorem
(Thm.~\ref{thm:loctrans}) there exists a 
unitary transformation $A_c$ acting just on $\alice$'s system $A$ such
that $(A_c \otimes I_B)\ket{\psi} = \ket{\psi_c}$, where $I_B$ is the
identity transformation on $\bob$'s system. Now the cheating $\alice$ (who's
intention is to reveal string $c$), applies the transformation $A_c$
to $\ket{\psi}$ and then continues with the rest of the reveal phase
as in the honest run. Let $\ket{\phi_c'}
\defeq \ket{\phi}$ be obtained from Thm.~\ref{thm:substate} and
hence, $\Tr |\ketbra{\phi_c} - \ketbra{\phi_c'}| \leq 2/\sqrt{r}$. Now
it can be seen that when $\bob$ makes the final checking $\POVM$, the
probability of success $\tp_c$ for $\alice$ is at least $(1 -
1/r)2^{-rk_c}(1 - 1/\sqrt{r})$ where $ k_c = \sD(\rho_c \| \rho_B) + 6
\sqrt{\sD(\rho_c \| \rho_B) + 1} + 4$. One way to see this is to imagine
that $\alice$ first measures the single qubit register $\C^2$ and then
proceeds with the rest of the reveal phase.  Now imagine that she
obtains one on this measurement which from Thm.~\ref{thm:substate} has
probability $(1 - 1/r)2^{-rk_c}$. Also once she obtains one, the
combined joint state of $\alice$ and $\bob$ is $\ket{\phi_c'}$ whose
trace distance with $\ket{\phi_c}$ is at most $2/\sqrt{r}$. Since
trace distance is preserved by unitary operations and is only smaller
for subsystems and since after this $\alice$ follows the rest of the
reveal phase honestly, we can conclude the following: the final state resulting
with $\bob$ will have trace distance at most $2/\sqrt{r}$ with the
state with him at the end of a completely honest run of the protocol
in which Alice starts with $c$ in the input register. Hence it follows from
Thm.~\ref{thm:dist} that $\bob$ will accept at the end with
probability at least $1 - 1/\sqrt{r}$ since he was accepting with
probability 1 in the complete honest run of the protocol . Hence the
overall cheating probability $\tp_c$ of $\alice$ is at least $(1 -
1/r)2^{-rk_c}(1 - 1/\sqrt{r})$.

Although here we have imagined $\alice$ doing an intermediate
measurement on the single qubit register $\C^2$, it is not necessary
and she will have the same cheating probability when she proceeds with
the rest of the honest protocol after just applying the cheating
transformation $A_c$ since the final qubits of $\bob$ will be in the same
state in either case. Now,
\begin{eqnarray*}
2^{a-n} & \geq & \sum_c p_c \tp_c  \\
& \geq & (1 - 1/r)(1 - 1/\sqrt{r}) \left( \sum_c  p_c 2^{-r(\sD(\rho_c
\| \rho_B) + 6 \sqrt{\sD(\rho_c \| \rho_B) + 1} + 4)} \right) \\ 
& \geq & (1 - 1/r)(1 - 1/\sqrt{r})2^{\sum_c -r  p_c(\sD(\rho_c \|
\rho_B) + 6 \sqrt{\sD(\rho_c \| \rho_B) + 1} +  4)} \\ 
& \geq &  (1 - 1/r)(1 - 1/\sqrt{r}) 2^{ -r(b + 6 \sqrt{b + 1} +  4) }
\end{eqnarray*}
The first inequality comes from definition of $a$ in
Defn.~\ref{def:QSC}. The third inequality comes from the convexity of
the exponential function and the fourth inequality comes from
definition of $b$ in Defn.~\ref{def:QSC}, Defn.~\ref{def:divinf} and
concavity of the square root function.

Now when $b > 15$, we let $r = 1 + \frac{1}{b}$ and therefore,
\begin{eqnarray*}
(1 - 1/r)(1 - 1/\sqrt{r}) 2^{ -r(b + 6 \sqrt{b + 1} +  4) }
& \geq & \frac{0.5}{(b + 1)^2} 2^{ -(b + 6 \sqrt{b + 1} +  7) } \\
& \geq & 2^{ -(b + 8 \sqrt{b + 1} +  8) }
\end{eqnarray*}

When $b \leq 15$, we let $r = 1 + 1/15$ and therefore,
\begin{eqnarray*}
(1 - 1/r)(1 - 1/\sqrt{r}) 2^{ -r(b + 6 \sqrt{b + 1} +  4) }
& \geq & 2^{ -(b + 6 \sqrt{b + 1} + 16) }
\end{eqnarray*}
Therefore we get always, $2^{a - n} \geq 2^{ -(b + 8 \sqrt{b + 1} +
16) } $ which finally implies,
$$ a + b + 8 \sqrt{b + 1} + 16  \geq  n .$$
\end{proofof}

\vspace{0.1cm}
\begin{proofof}{Thm.~\ref{thm:avgasym}}
Let $b_m$ represent the concealing parameter for $\Pi_m$.  It is easy to
verify from Lem.~\ref{lem:sadd} and definition of Holevo-$\chi$
information, Defn.~\ref{def:holevo}, that $b = b_m/m$.  Then
Thm.~\ref{thm:avgchi} when applied to $\Pi_m$ implies,
\begin{eqnarray*}
& \Rightarrow & a_m + b_m + 8 \sqrt{b_m + 2} + 17  \geq  mn \\
& \Rightarrow & \lim_{m \rightarrow \infty} \frac{1}{m}(a_m + b_m + 8 \sqrt{b_m +
2} + 17)  \geq  n \\
& \Rightarrow & a + b \geq n
\end{eqnarray*}
\end{proofof}

\vspace{0.1cm}
\noindent
\begin{proofof}{Thm.~\ref{thm:worst}} Let $\mu = \{\lambda_x\}$ be the
distribution on $\{0,1\}^n$ obtained from
Thm.~\ref{thm:remote}. Consider a cheating strategy of $\alice$ in
which she puts the superposition $\sum_x
\sqrt{\lambda_x} \ket{x}$ in the register where she keeps the commit
string. Let $c$ be the string in the cheating register of
$\alice$. Now by arguments as above probability of success $\tp_c$ for
$\alice$ is at least $(1 - 1/\sqrt{r})(1 - 1/r)2^{-rk_c}$ where $ k_c,
\rho_c, \rho$ being as before. Since for all $c, \sD(\rho_c \| \rho)
\leq b$ it implies (by setting $r$ appropriately) $\forall c,
\tp_c \geq 2^{-(b + 8 \sqrt{b + 1} + 16)}. $
\end{proofof}

\paragraph{Remark:} Let us now see how, with a good
probability overall, $\alice$ will be able to prevent herself from being
detected cheating by $\bob$. Let $\alice$ have $c$ in the cheating
register. Let $r_c$ be the probability of getting one on performing
the two outcome measurement (obtained from  Thm.~\ref{thm:substate})
after the commit phase as in the cheating strategy described above in
proof of Thm.~\ref{thm:avg}. In
case she gets one, she proceeds with the cheating strategy. In case
she gets zero, she tries to rollback so that she can successfully
reveal at least some string to $\bob$. For this she does the
following.

\begin{enumerate}

\item She applies the transformation $A_c^\dagger$ (that is inverse of $A_c$). 
\item She measures the input register in the computational basis and
say she obtains $x'$. 
\item She proceeds with the rest of the reveal phase as if her actual input was $x'$.

\end{enumerate}

Assume that $\alice$ obtains zero on performing the two-outcome
measurement as in the cheating strategy described above which happens
with probability $1 - r_c$. Now it can be verified that the trace
distance between $\ketbra{\psi_c}$ and the combined state of $\alice$ and $\bob$
after obtaining zero on performing the measurement is at most
$2r_c$. Since, $A_c^\dagger$ is unitary, this implies that the combined state of
$\alice$ and $\bob$ after applying $A_c^\dagger$,
and $\ketbra{\psi}$ will be at most $2r_c$. Now we can argue as before
that $\alice$ can reveal some string successfully to $\bob$ with
probability at least $1 - r_c$. Therefore overall, the probability
that $\alice$ will be able to reveal some string is at least $r_c + (1
- r_c)^2 \geq 1 - r_c$. Now since typically $r_c$ is quite small (like
$2^{-b}$), $1 - r_c$ is quite close to 1.

\subsection*{Acknowledgment} We thank Harry Buhrman, Matthias Christandl,
Hoi-Kwong Lo, Jaikumar Radhakrishnan, and Pranab Sen for
discussions. We also thank anonymous referees for suggestions
on an earlier draft.

\newcommand{\etalchar}[1]{$^{#1}$}

\appendix
\section{Separations for $\xi(\cE)$ and $\chi(\cE) $}
\label{sec:separation}
Let $\cE \defeq \{1/2^n, \rho_x \}$ be an ensemble with $x \in
\{0,1\}^n$. Let $\rho \defeq \sum_x 2^{-n} \rho_x$. Lets assume that for
all $x$, $\rho_x$ commutes with $\rho$ as is the case in classical
ensembles. We show that in this case $\xi(\cE) \geq \chi(\cE)$. Consider,
\begin{eqnarray*}
\xi(\cE) & = &  n + \log \sum_x \Tr (2^{-n} \rho^{-1/2} \rho_x)^2 \\
& = & \log \sum_x 2^{-n} \Tr (\rho^{-1/2} \rho_x)^2 \\
& \geq & 2^{-n} \sum_x \log \Tr  (\rho^{-1/2} \rho_x)^2  \mbox{\quad (from
concavity of $\log$ function)}\\
& = & 2^{-n} \sum_x \log \Tr (\rho_x \rho^{-1} \rho_x) \mbox{\quad (since
$\rho_x, \rho$ commute)}  \\
& \geq & 2^{-n} \sum_x \Tr \rho_x \log (\rho_x \rho^{-1}) \mbox{\quad (since
$\log \Tr BA \geq \Tr A \log B$, for $A,B$ quantum states)}  \\
& = & 2^{-n} \sum_x \Tr \rho_x (\log \rho_x - \log \rho ) \mbox{\quad (since
$\rho_x, \rho$ commute)}\\
& = & \chi(\cE)
\end{eqnarray*}

Next we show that there exists classical ensembles for which
$\xi(\cE)$ could be exponentially larger than $\chi(\cE)$. Consider the
ensemble of classical distributions $\{2^{-n}, P_x\}$ for $x \in
\{0,1\}^n$. Here each $P_x$ has support on $\{0,1\}^n$. Let
$\epsilon \in (0,1)$ be a constant. Let $P_x(x) = 2^{-\frac{\epsilon n}{2}}$
and let the other values for $P_x(y), y \neq x$ be the same. Let $P
\defeq \sum_x 2^{-n} P_x$. It is easy to verify that in this case $P$
is the uniform distribution on $\{0,1\}^n$. Now,

\begin{eqnarray*}
\xi(\cE) & = &  n + \log \sum_x \Tr (2^{-2n} P^{-1} P_x^2) \\
& = & -n + \log \sum_x \Tr (P^{-1} P_x^2) \\ & \geq & -n + \log \sum_x
2^{n( 1 - \epsilon) } \mbox{\quad (since for all $x$, $\Tr P^{-1}P_x^2
\geq 2^{n( 1 - \epsilon) }$ and since $\log$ is monotonic)}\\
& = & -n + \log 2^{n ( 2 - \epsilon)} \\
& = & n( 1 - \epsilon)
\end{eqnarray*}

Also we note that for all $x$, $\Tr P_x (\log P_x - \log P ) \leq
2^{-\frac{\epsilon n}{2}} \cdot n \cdot (1 -\epsilon/2)$ and hence,

\begin{eqnarray*}
\chi(\cE) & = & 2^{-n} \sum_x \Tr P_x (\log P_x - \log P ) \\
& \leq & 2^{-n} \sum_x 2^{-\frac{\epsilon n}{2}} \cdot n \cdot (1
-\epsilon/2) \\
& = & 2^{-\frac{\epsilon n}{2}} \cdot n \cdot (1 -\epsilon/2)
\end{eqnarray*}
Therefore by letting $\epsilon$ to be a constant very close to $0$, we can let
$\xi(\cE)$ to be very close to $n$ whereas $\chi(\cE)$ would still be
exponentially small in $n$.

\end{document}